\documentclass[%
 aip,
rsi,%
 amsmath,amssymb,
 reprint,%
]{revtex4-1}

\usepackage{amsmath}
\usepackage{amssymb}
\usepackage{amsfonts}
\usepackage{amssymb}
\usepackage[pdftex]{graphicx}
\usepackage{subfigure} 
\usepackage[english]{babel}
\usepackage{braket}
\usepackage{color}
\usepackage{mathtools}
\usepackage{threeparttable}
\usepackage[colorlinks,linkcolor=blue,anchorcolor=blue,citecolor=blue,urlcolor=blue]{hyperref}
\usepackage{babel}
\makeatother

\begin{document}

\preprint{APS/123-QED}

\title{Switching Purcell effect with nonlinear epsilon-near-zero media}

\author{Saman Jahani}
\affiliation{%
Department of Electrical and Computer Engineering,University of Alberta, Edmonton, AB, T6G 1H9 Canada.
}%
\affiliation{%
School of Electrical and Computer Engineering and Birck Nanotechnology Center, Purdue University, West Lafayette, IN, 47907 USA.
}%
\author{Hangqi Zhao}%
\affiliation{%
Department of Electrical and Computer Engineering,University of Alberta, Edmonton, AB, T6G 1H9 Canada.
}%
\author{Zubin Jacob}%
\email{zjacob@purdue.edu}
\affiliation{%
Department of Electrical and Computer Engineering,University of Alberta, Edmonton, AB, T6G 1H9 Canada.
}%
\affiliation{%
School of Electrical and Computer Engineering and Birck Nanotechnology Center, Purdue University, West Lafayette, IN, 47907 USA.
}%


\begin{abstract}
An optical topological transition is defined as the change in the photonic iso-frequency surface around epsilon-near-zero (ENZ) frequencies which can considerably change the spontaneous emission of a quantum emitter placed near a metamaterial slab. Here, we show that due to the strong Kerr nonlinearity at ENZ frequencies, a high-power pulse can induce a sudden transition in the topology of the iso-frequency dispersion curve, leading to a significant change in the transmission of propagating as well as evanescent waves through the metamaterial slab. This evanescent wave switch effect allows for the control of spontaneous emission through modulation of the Purcell effect. We develop a theory of the enhanced nonlinear response of ENZ media to {\it s} and {\it p} polarized inputs and show that this nonlinear effect is stronger for {\it p} polarization and is almost independent of the incident angle. We perform finite-difference time-domain (FDTD) simulations to demonstrate the transient response of the metamaterial slab to an ultrafast pulse  and fast switching of the Purcell effect at the sub-picosecond scale. The Purcell factor changes at ENZ by almost a factor of three which is an order of magnitude stronger than that away from ENZ. We also show that due to the inhomogeneous spatial field distribution inside the multilayer metal-dielectric super-lattice, a unique spatial topological transition metamaterial can be achieved by the control pulse induced nonlinearity. Our work can lead to ultra-fast control of quantum phenomena in ENZ metamaterials.
\end{abstract}

\maketitle


{

Progress in quantum information processing at optical frequencies can be accelerated by developments in the control and manipulation of light-matter interaction at the nanoscale \cite{lodahl_interfacing_2015}. Some recent achievements include on-chip superconducting single photon detectors \cite{schuck_waveguide_2013}, spin routing of single photons \cite{mitsch_quantum_2014,mechelen_universal_2016,kalhor_universal_2016} long range dipole-dipole interactions \cite{cortes_super-coulombic_2017,gonzalez-tudela_subwavelength_2015}, enhancement or suppression of vacuum fluctuations in a selected spectrum \cite{yablonovitch_inhibited_1987, krishnamoorthy_topological_2012, vahala_optical_2003, guo_fluctuational_2014,gonzalez-tudela_subwavelength_2015,aoki_observation_2006, reithmaier_strong_2004,srinivasan_linear_2007,yoshie_vacuum_2004,choy_enhanced_2011, faraon_resonant_2011, akimov_generation_2007}, modulation and switching of light sources \cite{miller_are_2010,gibbs_optical_2012, yanik_all-optical_2003, krasnok_all-optical_2018}, control of evanescent waves \cite{jahani_transparent_2014,jahani_controlling_2018}, and switching at the single photon level \cite{bermel_single-photon_2006, chang_single-photon_2007, tiarks_single-photon_2014, gorniaczyk_single-photon_2014}. However, all-optical manipulation and switching at the same time is a difficult task as light hardly interacts with light, and it requires carefully engineered nanostructures to confine light with strong nonlinear functionality.

Structures with epsilon-near-zero (ENZ) response have emerged as a unique nanophotonic platform for controlling light because of field enhancement and spatially coherent wave behavior in these media \cite{liberal_near-zero_2017,jahani_all-dielectric_2016, moitra_realization_2013,reshef_direct_2017}. This has led to many interesting phenomena and applications such as controlling thermal emission \cite{molesky_high_2013}, observation of Ferrel-Berreman modes \cite{newman_ferrellberreman_2015}, super-coupling \cite{silveirinha_theory_2007}, and enhanced dipole-dipole interaction \cite{cortes_super-coulombic_2017,mahmoud_dipole-dipole_2017}. Recently, strong nonlinearity in ENZ media has become a domain of interest because of significant change in the refractive index around the ENZ frequencies as well as the field enhancement due to boundary effects at the high contrast interfaces \cite{alam_large_2016,kinsey_epsilon-near-zero_2015,caspani_enhanced_2016}, phase mismatch-free propagation \cite{suchowski_phase_2013}, and strong nonlocality at ENZ \cite{wurtz_designed_2011,nicholls_ultrafast_2017}. 

In this paper, we propose a hyperbolic metamaterial (HMM) slab operating around ENZ frequencies, in which a nonlinear gate signal controls and switches the decay rate of light emitted by a dipole emitter above the slab (Fig.~\ref{fig:fig1HMM}.(a)). This can give rise to a modulation approach for spontaneous emission through switching the Purcell effect. First, we develop a theory of nonlinear light propagation in anisotropic ENZ media for {\it s} and  {\it p} polarized light.  We report on the formation of a unique ENZ-transition metamaterial induced by nonlinear wave interactions and inhomogeneity of the field distribution inside the medium. As a result, the spontaneous decay rate of a quantum emitter close to the slab is considerably changed. We propose experiments to verify our predictions taking into account the role of dispersion, absorption, and finite unit-cell size in practical metal-dielectric nonlinear HMMs. Furthermore, to understand the response of the metamaterial to ultra-fast pulses, we carry out finite-difference time-domain (FDTD) simulations. We propose optimum gate pulse widths to open up the possibility for switching Purcell effect with sub-picosecond response times using ENZ media. 

\begin{figure*}[htbp]
\centering
\begin{tabular}{cc}

\includegraphics{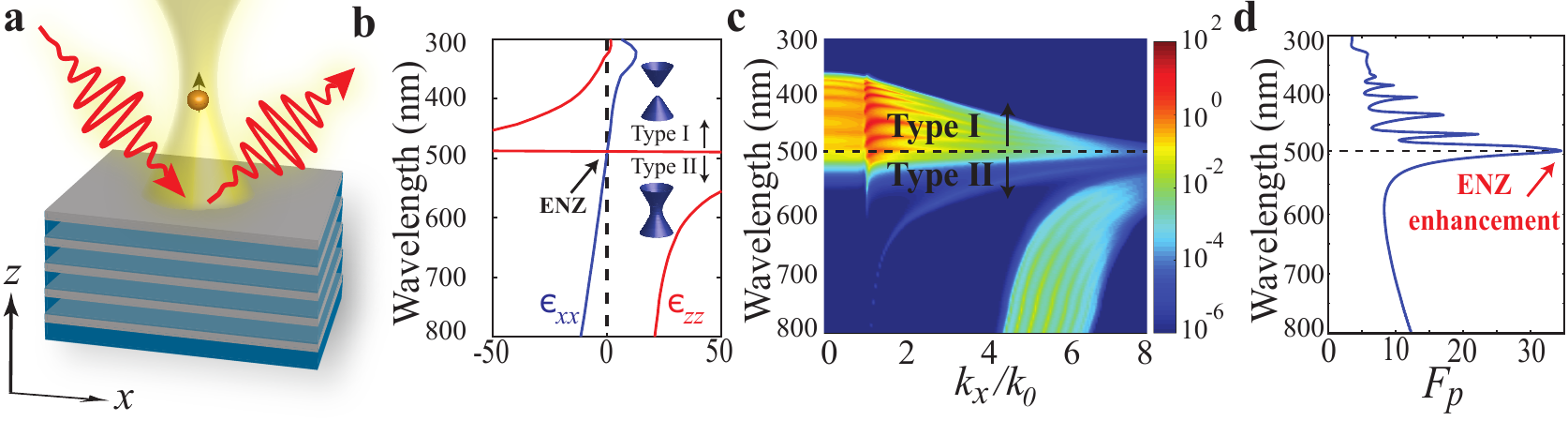}

\end{tabular}
\caption{ {\bf Switching of Purcell effect.} (a) A nonlinear gate signal controls the transmission of evanescent waves emitted by a quantum emitter above HMM around the ENZ frequencies. (b) The real part of effective permittivity of the Ag/TiO$_2$ multilayer metamaterial with Silver filling fraction of $\rho=0.5$. The inset illustrates the iso-frequency dispersion curve of type I and type II HMMs. An optical topological transition is observed at ENZ. (c)  Transmission of propagating and evanescent p-polarized waves through an Ag/TiO$_{2}$ multilayer with a total thickness of 400 nm and a periodicity of $\Lambda=50$~nm. The transmission of high-k modes in both regions is high, but in type I region, the effect is stronger, especially close to the ENZ wavelengths. Due to the finite thickness of the unit-cell, the ENZ wavelength is red-shifted in multilayer structure in comparison with EMT. (d) Purcell factor for a vertically oriented dipole above the slab modeled by EMT. A rapid change in the Purcell factor is observable at ENZ due to the optical topological transition. Other peaks in type I correspond to slow-light modes of the HMM slab.}
\label{fig:fig1HMM}
\end{figure*}

{

HMMs are a class of non-magnetic uniaxial metamaterials with double-sheeted (type I) or single-sheeted (type II) hyperbolic iso-frequency dispersion curves \cite{cortes_quantum_2012, poddubny_hyperbolic_2013, shekhar_hyperbolic_2014}. There are two well-known approaches for practical realization of HMMs: metallic  nanowires in a dielectric host matrix \cite{yao_optical_2008, kabashin_plasmonic_2009, starko-bowes_optical_2015}, and metal-dielectric multilayers \cite{newman_ferrellberreman_2015, liu_far-field_2007}. Here, we emphasize the second approach due to ease of fabrication and experimental verification. Figure.~\ref{fig:fig1HMM}.(b) displays the parallel ($\epsilon_{xx}=\epsilon_{yy}$) and normal ($\epsilon_{zz}$) components of the effective dielectric tensor of an Ag/TiO$_2$ multilayer metamaterial derived from Maxwell-Garnett effective medium theory (EMT) \cite{milton_theory_2002}. An optical topological transition from type I to type II is seen at the ENZ wavelength. HMMs show  interesting characteristics due to their open dispersion surfaces and the coupling of high-k modes (evanescent waves in vacuum) to the metamaterial modes \cite{cortes_quantum_2012}.  Figure~\ref{fig:fig1HMM}.(c) shows the transmission of low-k ($k_x<k_0$) and high-k ($k_x>k_0$) modes through the multilayer using transfer matrix method. The peaks correspond to the coupling of evanescent waves to the metamaterial modes with large wave vectors. We emphasize that even though the linear properties of HMMs have been studied extensively, the ``evanescent wave switch" we introduce in the paper using nonlinear interactions has not been studied till date. 

If a quantum emitter is placed near an HMM slab, the non-radiative evanescent waves of emitter are coupled to the high-k modes of metamaterials. As a result, spontaneous emission rate is enhanced \cite{jacob_broadband_2012,kidwai_dipole_2011, shekhar_momentum-resolved_2017}. Figure~\ref{fig:fig1HMM}.(d) displays the spontaneous emission rate spectrum of a quantum emitter above the multilayer structure modeled by EMT. The dipole emitter is assumed vertically polarized which is placed 20 nm above the metamaterial slab, and the emission rate is normalized to the emission of emitter in free space. This enhancement factor is known as Purcell factor ($F_p$) \cite{cortes_quantum_2012}. Enhanced Purcell effect is obtained in both regions, but larger enhancement is observed in type I region due to the high transmission of low-k as well as high-k modes. The strongest peak around $\lambda=500$~nm corresponds to the coupling of power to the resonating mode of the slab when ENZ response is achieved along the direction perpendicular to the layers. Other peaks in type I region correspond to slow-light modes of HMMs \cite{cortes_photonic_2013}. We now show that the strong nonlinearity at ENZ can help to effectively change the type of HMMs. This opens up the possibility of switching the Purcell effect.

\begin{figure}[htbp]
\centering
\begin{tabular}{cc}
\includegraphics{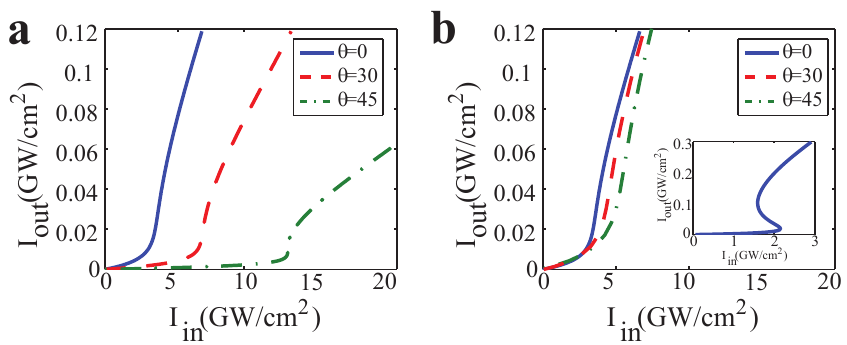}
\end{tabular}
\caption{{\bf Nonlinear response at ENZ.} Nonlinear response of (a) {\it s} polarized and (b) {\it p} polarized incidences at different incident angles for the multilayer structures in Fig.~\ref{fig:fig1HMM} at $\lambda=522$~nm which is slightly longer than the ENZ wavelength.  As the input power increases, the transmission increases due to the change in effective permittivity of the multilayer slab. The power level is independent of the incident angle for the p-polarization. It is important to note that ENZ media show bistable behavior in the presence of low loss (inset). This bistability does not exist if we take into account nonlinear absorption i.e.imaginary part of ${\chi }^{(3)}$.}
\label{fig:fig2HMM}
\end{figure}

\begin{figure}[ht]
\centering
\begin{tabular}{cc}
\includegraphics{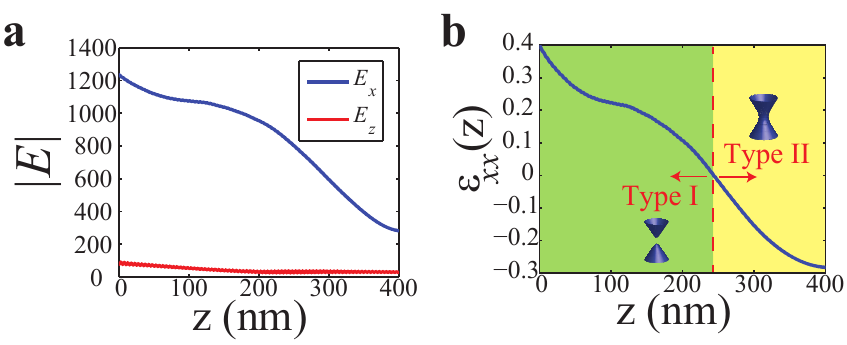}
\end{tabular}
\caption{{\bf Optically induced ENZ transition metamaterial.} (a) Electric field distribution for the {\it p} polarized incidence when I${}_{in}$=8 GW/cm${}^{2}$ and the incident angle is 45$^{\circ}$. Since the operating wavelength is close to the ENZ and epsilon-near-pole (ENP) resonance, $\epsilon_z$  is very large which causes the electric field E${}_{z}$ to become very small in comparison with E${}_{x}$. (b) Nonlinear ENZ transition metamaterials: The nonlinear effective permittivity of the HMM slab when the pump is on. We can see the topological transition from type II to type I occurs as a function of distance (as opposed to wavelength) due to the inhomogeneous power distribution inside the multilayer slab. This shows a unique regime of the multi-layer super-lattice functioning as a transition metamaterial.}
\label{fig:fig3HMM}
\end{figure}

Metals have very large Kerr nonlinearity (${\epsilon }_{NL}={\epsilon }_L+12\pi {\chi }^{(3)}{\left|E\right|}^2$ ~\cite{boyd_nonlinear_2003, kauranen_nonlinear_2012} where ${\epsilon }_L$, ${\chi }^{(3)}$, and $\left|E\right|$ are linear permittivity, Kerr susceptibility, and the electric field amplitude, respectively) which is orders of magnitude larger than that of dielectrics \cite{kauranen_nonlinear_2012}. However, since they are strongly reflective at optical frequencies, the electric field decays very fast inside the bulk metals, so their nonlinearity is not easily accessible \cite{bennink_accessing_1999, sarychev_electromagnetic_2000, lepeshkin_enhanced_2004}. Enhancing the nonlinear response of metals in inhomogeneous structures at transparent window has been proposed for applications such as optical limiting \cite{scalora_nonlinear_2006} and switching of propagating waves \cite{husakou_steplike_2007,campione_optical_2015}. As illustrated in Fig.~\ref{fig:fig1HMM}.(c), metal/dielectric multilayer structures are transparent in HMM type I region where ${\epsilon }_{xx}>0$. Thus, the nonlinear response in that range of frequencies can considerably be enhanced. Furthermore, a small change in permittivity around the ENZ point can lead to a topological transition in HMMs from type II to type I.

To calculate the nonlinear response, we use a finite difference method to discretize the wave equation in structures which are inhomogeneous only in one direction i.e. stratified structures \cite{bennink_accessing_1999}.  Nonlinear wave propagation within a stratified medium has been analytically studied only for {\it s} polarized incidences \cite{bennink_accessing_1999} and has been approximated for {\it p} polarized incidences \cite{argyropoulos_negative_2013,shramkova_nonreciprocal_2017,campione_optical_2015}. However, the recent observation of strong nonlinearity at the ENZ wavelengths for {\it p} polarized incidences \cite{alam_large_2016} raises the need for a robust analytical formulation to study full-wave nonlinear wave propagation for {\it p} polarization as well. We have modified the method proposed by \citealt{bennink_accessing_1999} to solve the nonlinear response for {\it p} polarized waves. 

For an {\it s} polarized plane wave propagating through a multilayer structure which is inhomogeneous only in the $z$ direction, the wave equation can be written as: ${d^2E_y\left(z\right)}/{{dz}^2}+\left[{k_0}^2{\epsilon }_{NL}\left(z\right)-{k_x}^2\right]E_y\left(z\right)=0$. This equation can be solved backward by applying appropriate boundary conditions \cite{bennink_accessing_1999}. The nonlinear response of the output power for the  multilayer structure is plotted in Fig.~\ref{fig:fig2HMM}.(a) as a function of the input power. We assume ${\chi }^{(3)}$ is $2.49\times {10}^{-8}+i7.16\times {10}^{-9}$ esu for silver \cite{husakou_steplike_2007} and it is negligible for TiO${}_{2}$. As the input power increases, the transmission increases slowly up to a point and then the transmission soars rapidly. At this threshold, the electric field enhancement effectively flips the sign of ${\epsilon }_{NL}$ from negative to positive. Thus, the structure behaves like a type I HMM. As the incident angle increases, we need higher input power to pass the threshold.
If we ignore the imaginary part of ${\chi }^{(3)}$, which corresponds to the two-photon absorption \cite{boyd_nonlinear_2003}, the nonlinear response at the ENZ threshold is bistable: there are two stable results for the same input (Fig.~\ref{fig:fig2HMM}.(b) inset). But the bistability does not occur in practice if we take the two-photon absorption into account.

The wave equation for {\it p} polarized incidences is more complicated and can be written as (See Supplemental Materials):
\begin{widetext}
\begin{subequations}\label{GrindEQ__3_} 
\begin{align}
\frac{d^2E_x\left(z\right)}{{dz}^2}+\left({k_0}^2{\epsilon }_{NL}\left(z\right)-{k_x}^2\right)\frac{{k_x}^2}{{k_0}^2{\epsilon }_{NL}\left(z\right)}\frac{d}{dz}\left[\frac{1}{{k_0}^2{\epsilon }_{NL}\left(z\right)-{k_x}^2}\right]\frac{dE_x\left(z\right)}{dz}+\left({k_0}^2{\epsilon }_{NL}\left(z\right)-{k_x}^2\right)E_x\left(z\right)=0, \\
E_z=\frac{ik_x}{{k_0}^2{\epsilon }_{NL}\left(z\right)-{k_x}^2}\frac{dE_x\left(z\right)}{dz}. 
\end{align} 
\end{subequations}
\end{widetext}
 
Since ${\epsilon }_{NL}$ depends on both $E_x$ and $E_z$, the wave equation for the {\it p} polarized incidence cannot be solved explicitly as what we did for the {\it s} polarization case. However, with a recursive algorithm, accurate result is achievable. The nonlinear response for the {\it p} polarized case is plotted in Fig.~\ref{fig:fig2HMM}.(b). It is seen that the response is almost independent of the incident angle, contrary to the {\it s} polarized cases.

Figure~\ref{fig:fig3HMM}.(a) displays the electric field distribution inside the EMT structure at $\lambda=500$~nm when $I_{in}=8$~GW/cm$^2$. Since the transmission is not significantly high, the electric field intensity is larger around the input interface. Due to the inhomogeneous field distribution, the permittivity changes inhomogeneously inside the structure. Hence, we can see a transition point at which the effective permittivity sign is flipped (Fig.~\ref{fig:fig3HMM}.(b)). This type of inhomogeneous metamaterials in which the iso-frequency dispersion curve changes from type II to type I HMMs is known as ``transition ENZ metamaterials'' \cite{litchinitser_metamaterials:_2008, dalarsson_analytical_2009, wells_nonlocal_2017}. This phase change is controlled by the electromagnetic field and is fundamentally different from the natural phase change materials like VO$_{2}$ \cite{mott_metal-insulator_1974}. Note that the operating wavelength of the structure is tunable by changing layers thickness or changing the layer materials. 

{
\begin{figure}[htbp]
\centering
\begin{tabular}{cc}

\includegraphics{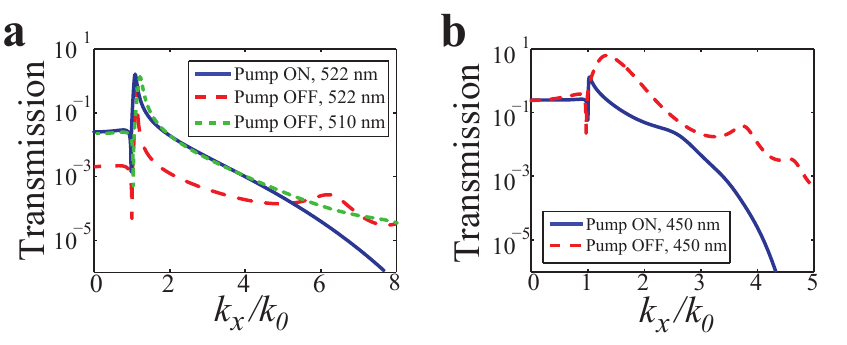}

\end{tabular}
\caption{{\bf Effect of nonlinear gate pulse on the linear transmission (a) around and away from the ENZ point.} (a) Transmission of signal through the multilayer medium in the presence of the pump in comparison with that when the pump is off at $\lambda=522$~nm. The nonlinearity allows us to suppress or enhance transmission of evanescent waves. The nonlinear pulse suppresses the Purcell factor from 41 to around 15. (b) Comparison of the transmission away from the transition point. The nonlinear pulse does not change the transmission for a wide range of wave-vectors. Thus the pump signal does not change the Purcell factor considerably.}
\label{fig:fig4HMM}
\end{figure}

\begin{figure}[htbp]
\centering
\begin{tabular}{cc}

\includegraphics{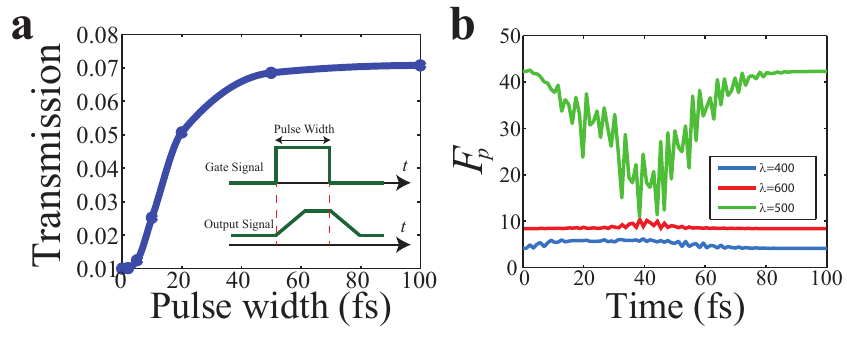}

\end{tabular}
\caption{{\bf Nonlinear response of the HMM slab to a finite pulse width.} (a) Nonlinear transmission versus the pulse width at $\lambda=500$ nm using FDTD method. The electric field amplitude is 10$^4$ V/m. When the pulse-width is very short, the electric field cannot be distributed in the entire space of the slab, so the pump cannot increase the transmission considerably. As the pulse-width increases, the transmission increases, and it approaches to the steady-state response when the pulse width is longer than 50~fs. (b) The Purcell factor at $\lambda=500$ nm versus time in the presence of the gate signal. The nonlinear gate signal can switch the Purcell effect. We can see that the change in Purcell factor far away from the ENZ wavelength is negligible.}
\label{fig:fig5HMM}
\end{figure}

Propagation of a nonlinear light passing through a multilayer metamaterial slab not only modifies the transmission of the propagating waves, but also changes the tunneling of the evanescent waves through the metamaterial slab. This allows us to control spontaneous emission rate of an emitter near the slab. Figure~\ref{fig:fig4HMM} compares transmission through the slab with and without the presence of the nonlinear signal near ($\lambda=$522 nm) and away ($\lambda=$450 nm) from the ENZ wavelength. The control signal is a {\it p} polarized wave at $\lambda=522$ nm with $I_{in}=8$ GW/cm$^2$ and $\theta =45^{\circ}$. An observable change in transmission of both propagating and evanescent waves is seen around the ENZ, and it looks similar to the transmission at a shorter wavelength ($\lambda=510$~nm) when the pump signal is off. This shows how the nonlinear response due to the control pulse effectively shifts the dispersion of the metamaterial slab around the ENZ. Away from the ENZ, the control pulse does not change the topology of the dispersion curve and hence the change in transmission is not significant for a wide range of wave vectors (Fig.~\ref{fig:fig4HMM}.(b)). 
For the case at ENZ, the gate control signal changes the Purcell factor from 41 to around 15. This observable change allows us to control the coupling of the emitted power by a quantum source to the high-k modes of HMMs.
 
To demonstrate the transient response of the switching process, we have simulated the nonlinear response of the metamaterial in presence of a gate signal with a finite pulse width using a commercial FDTD software \cite{lumerical_fdtd_solutions}. For simplicity, we consider homogenized metamaterial modeled by EMT, and we ignore two photon absorption and the dispersion of $\chi^{(3)}$. The input signal is {\it p} polarized and $I_{in}=8$ GW/cm$^2$. The central wavelength of the pulse is at $\lambda=500$ nm. Figure~\ref{fig:fig5HMM}.(a) shows the transmission of a normal incident wave as a function of the pulse width. If the pulse width is very short, the electric field cannot influence the entire structure. Thus, the change in the dielectric constant and the nonlinear effect in the transmission is not considerable. Thus, the rise time of the nonlinear response is limited by the material response and the velocity of wave propagation through the HMM slab. As we increase the pulse-width, the transmission increases gradually (Fig.~\ref{fig:fig5HMM}.(a) inset) and  the steady state response is reached when the pulse width is longer than 50 fs. It means that 50 fs is enough for the wave to pass and be distributed in the metamaterial slab and simultaneously change the nonlinear permittivity of the slab. Note that if we consider $\chi^{(3)}$ to be dispersive, the nonlinear dielectric response is not instantaneous, but the response to the nonlinear pulse can still occur at sub-picosecond scale \cite{stockman_nanoplasmonics:_2011,alam_large_2016,kinsey_epsilon-near-zero_2015}. 

Figure~\ref{fig:fig5HMM}.(b) shows the switching of the Purcell effect as a function of time when the pulse width is 50 fs using the time-dependent nonlinear permittivity derived from the FDTD simulations. More than two-fold Purcell factor reduction is seen near the transition point when the gate pulse is on. The numerical simulations are in good agreement with the analytical calculations. The change in the Purcell factor far away from the transition point is less than 10\%. The change in Purcell factor for isotropic media at ENZ is also negligible \cite{chebykin_strong_2015} (See Supplemental Materials). This means that switching of the Purcell effect is only possible around the ENZ wavelength.

{

In conclusion, we have proposed an all-optical nonlinear approach to control the transmission of propagating waves as well as the tunneling of the evanescent waves through an HMM slab around ENZ wavelengths. We have shown that the nonlinear effect is strong enough to change the topology of the HMM slab. This can lead to a significant change in the Purcell factor of a quantum emitter near the slab. We have performed full-wave FDTD simulations to show the effect of pulse-width on the nonlinear response of the slab. The switching can happen at sub-picosecond time scales. Our predicted nonlinear effect is spectrally sensitive and can lead to control of quantum and nonlinear photonic phenomena in the ENZ regime.

This material is based upon work supported by the U.S. Department of Energy, Office of Basic Energy Sciences under award number  \# DE-SC0017717.

\bibliography{Evanescentwavetransistor}

\end{document}